\documentclass[lettersize,journal]{IEEEtran}
\usepackage{amsmath,amsfonts}
\usepackage{algorithmic}
\usepackage{algorithm}
\usepackage{array}
\usepackage{amssymb,color,amsbsy,bm}
\usepackage[caption=false,font=normalsize,labelfont=sf,textfont=sf]{subfig}
\usepackage{multirow}
\usepackage{float}
\usepackage{amsthm}
\newtheorem{theorem}{Theorem}
\usepackage{textcomp}
\usepackage{stfloats}
\usepackage{url}
\usepackage{verbatim}
\usepackage{graphicx}
\usepackage{cite}
\newtheorem{lemma}[theorem]{Lemma}
\allowdisplaybreaks
\usepackage{color,soul}

\usepackage{textcomp}
\usepackage{xcolor}
\begin{document}

%\title{ISO and DSO Coordination: A Parametric Programming Approach}
\title{Transmission and Distribution Coordination for DER-rich Energy Markets: A Parametric Programming Approach}

\author{Mohammad~Mousavi,~\IEEEmembership{Student~Member,~IEEE},~Meng~Wu,~\IEEEmembership{Member,~IEEE}
        % <-this % stops a space
\thanks{The authors are with the School of Electrical, Computer and Energy Engineering, Arizona State University, Tempe, AZ 85281 USA  (e-mail: mmousav1@asu.edu; mwu@asu.edu).}}% <-this % stops a space
%\thanks{Manuscript received April 19, 2021; revised August 16, 2021.}}

\markboth{Journal of \LaTeX\ Class Files,~Vol.~14, No.~8, August~2021}%
{Shell \MakeLowercase{\textit{et al.}}: A Sample Article Using IEEEtran.cls for IEEE Journals}


\maketitle

\begin{abstract}
In this paper, a framework is proposed to coordinate the operation of the independent system operator (ISO) and distribution system operator (DSO). The framework is compatible with current practice of the U.S. wholesale market  to enable massive distributed energy resources (DERs) to participate in the wholesale market. The DSO builds a bid-in cost function to be submitted to the ISO market through parametric programming. Once the ISO clears the wholesale market, the dispatch and payment of the DSO are determined by ISO. Then, the DSO determines the dispatch and payment of the DER aggregators. To compare the proposed framework, an ideal case is defined in which DER aggregators can participate in the wholesale market directly and ISO overseas operation of both transmission and distribution systems. We proved 1) the dispatches of the proposed ISO-DSO coordination framework are identical to those of the ideal case; 2) the payments to each DER aggregator are identical in the proposed framework and in the ideal case. Case studies are performed on a small illustrative example as well as a large test system which includes IEEE 118 bus transmission system and two distribution systems - the IEEE 33 node and IEEE 240 node test systems.
\end{abstract}

\section{Introduction}
\IEEEPARstart{T}{he} Federal Energy Regulatory Commission (FERC) issued Order No. 2222 which requires all the US independent system operators (ISOs) to open their wholesale energy and ancillary service markets to the distributed energy resources (DER) aggregators market participation \cite{ferc_2222}. However, the uncontrolled participation of DER aggregators in the wholesale market may cause security and reliability issues in the distribution system. To overcome this issue, designing a distribution system operator (DSO) for coordinating the DER aggregators has been proposed \cite{mousavi2021dso}. However, the operation of the DSO should be compatible with the current practice of the wholesale markets operated by independent system operators (ISOs). Hence, the operation of the DSO and ISO should be coordinated. %The coordination of the DSO and ISO is an important topic in the wholesale market and is ongoing research \cite{sun2021research,wu2020future}.

Recently, several works have studied the coordination of the ISO and DSO  \cite{chen2021distribution,hassan2018energy,jiang2022flexibility,steriotis2022co,zhang2022optimal,el2021coordinated,moret2020loss,renani2017optimal, wang2021real,yin2021stochastic,haider2020toward,bragin2021tso,khodadadi2020non,yuan2017hierarchical,lin2017decentralized,zhou2020economic,jiang2022risk,liu2020integrating,liu2020accurate,lin2019determination,guo2016multi}. They fall into three categories based on the modeling and solution method. The first category proposed bi-level models with the ISO and DSO markets modeled at two levels. The problem is transformed to single level optimization \cite{chen2021distribution,hassan2018energy,jiang2022flexibility,steriotis2022co,zhang2022optimal,el2021coordinated}. In \cite{chen2021distribution}, a bi-level optimization model is proposed for DSO market clearing and pricing considering ISO-DSO coordination. The clearing conditions for the DSO and ISO markets are proposed in the upper-level and lower-level problems, respectively. The problem is converted to mixed-integer linear programming via an equilibrium problem with equilibrium constraints (EPEC) approach. In \cite{hassan2018energy}, a bi-level optimization model is proposed for the energy storage sizing and siting problem in the DSO-ISO coordination framework. In \cite{jiang2022flexibility}, a bi-level optimization model is proposed for the energy and flexibility market in which, the upper-level models the clearing conditions of the transmission level market while, in the lower level, clearing conditions of the distribution level market are modeled. In \cite{steriotis2022co}, a coordination scheme is proposed for energy service providers, transmission system operator (TSO), and DSO for DER planning while coordinating the operation of the TSO and DSO based on bi-level optimization. In the upper-level problem, the DSO cost is minimized and the profit of the energy service providers is ensured while the lower-level problem models the transmission level constraints and wholesale market. In \cite{zhang2022optimal}, a bi-level optimization model is proposed for the coordinated operation of active distribution networks with multiple virtual power plants in joint energy and reserve markets operated by the DSO. At the upper level, the DSO minimizes the total operational cost of the distribution system while in the lower level problem, virtual power plants maximize their profit. In \cite{el2021coordinated} a tri-level coordinated scheme for transmission and distribution (T\&D) systems expansion planning is proposed. In the first and second levels, the expansion planning of T\&D systems operated by the TSO and DSO are proposed, respectively. The third level is the economic dispatch problem performed by the ISO. The multi-parametric programming approach is used to convert the multi-level optimization problem into a single-level optimization problem.  Bi-level optimization models are computationally expensive and hard to solve especially for large systems. These approaches place a high computation burden on the wholesale market and is not compatible with the current practice of the wholesale market. 

The second category of works uses decentralized models and some of them use decomposition algorithms to decouple the ISO and DSO markets \cite{moret2020loss,renani2017optimal,wang2021real,yin2021stochastic,haider2020toward,bragin2021tso,khodadadi2020non,yuan2017hierarchical,lin2017decentralized,zhou2020economic,jiang2022risk}. In \cite{moret2020loss}, an extension of the decentralized market framework is proposed to consider loss allocation and its impact on the market outcome. However, the decentralized market framework is not compatible with the current market structures. In \cite{renani2017optimal}, a transactive market framework starting from the ISO to the DSO is proposed. The DSO runs the transactive market using an iterative method. However, the convergence of the proposed method is not guaranteed. In \cite{wang2021real}, a Nash bargaining-based method is proposed for the market-clearing process and the ISO-DSO coordination. The proposed model requires high ISO-DSO communication burden within each wholesale market clearing interval. In \cite{yin2021stochastic}, a three-stage unit commitment (UC) is proposed for transmission-distribution coordination based on stochastic programming. A convex AC branch flow model is proposed to handle the distribution grid's physical constraints. However, stochastic programming is not compatible with the current practice of the wholesale market.
 In \cite{haider2020toward}, a distributed optimization algorithm is proposed for modeling the DSO retail market considering energy and ancillary services. However, the DSO's impact on wholesale market clearing is not considered. In \cite{bragin2021tso}, the optimal operation and coordination of the ISO-DSO are proposed. A decomposition algorithm is proposed and the original problem is decomposed into ISO and DSO sub-problems. In \cite{khodadadi2020non}, a non-cooperative game approach is proposed for ISO-DSO coordination in which they optimize their operational costs. The approaches in \cite{bragin2021tso, khodadadi2020non} are hard to solve for large systems. In \cite{yuan2017hierarchical}, a coordination framework for coordinating the economic dispatch of the TSO and DSO is proposed. Benders' decomposition is used for solving the proposed problem. In \cite{lin2017decentralized}, a coordination framework is proposed for the dynamic economic dispatch problem of the ISO and DSO. A decentralized approach is proposed to solve this problem. Nevertheless, References \cite{yuan2017hierarchical,lin2017decentralized} do not propose any market framework or settlement. In \cite{zhou2020economic}, an economic dispatch for co-optimization of T\&D systems is proposed. Primal-dual gradient algorithm based on the Lagrangian function is proposed to solve the co-optimization problem. However, the proposed method is not appropriate for a large number of DERs in the distribution system as it places so much computation burden on the economic dispatch of the wholesale market. %In \cite{bagheri2022igdt}, a decentralized method is proposed for the coordination of the TSO-DSO wholesale and retail markets based on optimality condition decomposition. The proposed method is equipped with two-stage for handling uncertainties. However, the convergence of the proposed problem is not guaranteed. In \cite{jiang2022risk}, a stochastic programming risk-averse approach is proposed for TSO-DSO coordinated distributed dispatching. The proposed method tries to minimize electricity purchasing costs and potential risk costs by proposing two optimization models solved by a distributed algorithm. The proposed method does not guarantee convergence and also is computationally expensive.

The third group of works proposed equivalent models for T\&D coordination \cite{liu2020integrating,liu2020accurate,lin2019determination,guo2016multi}. In \cite{liu2020integrating}, a feasible region-based approach is proposed for the integration of DERs into the wholesale market considering the physical constraints of the distribution system operated by the DSO. In \cite{liu2020accurate}, a multi-port power exchange model is proposed to integrate the high penetration of the DERs into the wholesale market considering the physical constraints of the distribution network. The approaches in \cite{liu2020integrating} and \cite{liu2020accurate} require modeling a transformed version of the distribution level constraints in the ISO market clearing problem, which significantly increases the modeling and computational complexity for the ISO. 
%In \cite{liu2020accurate}, a multi-port power exchange model is proposed to integrate the high penetration of the DERs into the wholesale market considering the physical constraints of the distribution network. 
%It does not propose any market for the DSO and subsequently does not propose any market settlement for the DSO market. 
In \cite{lin2019determination}, a unified equivalent model for external power networks based on multi-parametric programming is proposed for determining the transfer capacity of tie lines. However, they have not considered the distribution system and the market settlement of these external power networks. In \cite{guo2016multi}, a coordinated economic dispatch is proposed for a multi-area power system based on parametric programming. However, no market framework is proposed. Besides, this approach requires iterative communications beteween the coordinator and each economic dispatch sub-area before reaching convergence for the overall coordinated problem. This places very high communication burden between the market operators which is difficult to be implemented in real world applications.

Ideally speaking, the ISO-DSO coordination for DER integration in the wholesale market should satisfy the following requirements: 1) There should be no exchange of grid models between T\&D systems, in order to eliminate data confidentiality/privacy issues and avoid additional modeling/computational burden for ISO or DSO. 2) The coordination procedure should introduce no or minimal change to existing ISO wholesale market clearing procedure. 3) The coordination procedure should minimize the communication burden between ISO and DSO, by exchanging the minimal amount of public data and also by avoiding iterative T\&D communications within each wholesale market clearing interval.

So far, there is no existing ISO-DSO coordination which fully satisfies the above requirements. Existing works either 1) exchange T\&D grid models \cite{chen2021distribution,hassan2018energy,jiang2022flexibility,steriotis2022co,zhang2022optimal,el2021coordinated}; 2) introduce significant changes to existing ISO market clearing \cite{moret2020loss,renani2017optimal,yin2021stochastic,bragin2021tso,khodadadi2020non,yuan2017hierarchical,lin2017decentralized,zhou2020economic}; or 3) introduce high ISO-DSO communication burden and iterative ISO-DSO communications within each wholesale market clearing interval \cite{wang2021real, haider2020toward,zhou2020economic,jiang2022risk,liu2020integrating,liu2020accurate}.

This paper proposes an ISO-DSO coordination framework which satisfies all the above requirements. The proposed framework coordinates the operation of ISO and DSO to leverage the wholesale market participation of DER aggregators while ensuring the secure operation of distribution grids. The proposed coordination framework is based on parametric programming. The DSO builds the bid-in cost function based on the distribution system market considering its market participants' constraints and distribution system physical constraints including the power balance equations and voltage limitation constraints.  The DSO submits the resulting bid-in cost function to the wholesale market operated by the ISO. After the clearance of the wholesale market, the DSO determines the share of its DSO market participants (i.e., aggregators). Case studies are performed to verify the effectiveness of the proposed method. 

%Still, there are a lot of questions that remain unexplored. How should a DSO determine its bid into the wholesale market? How should the DSO distribute the won share in the wholesale market to the retail market participants? What is a coordination framework that is compatible with the current structure of the wholesale market? How does the computation time of the ISO problem changes when DSO takes responsibility for the DER market participation in its territory? 

%The above questions are answered in this work. 

This paper extends our prior works in \cite{mousavi2021dso,mousavi2020dso,mousavi2020two,mousavi2022iso}. To the best of our knowledge, this is the first ISO-DSO coordination framework which fully satisfies all the above performance requirements for ideal and practical ISO-DSO coordination. This is achieved by the following major contributions:

%This \blue{work satisfies the above performance requirements for ideal ISO-DSO coordination by} the following major contributions:

 \begin{itemize}
	\item A framework is proposed to coordinate the operation of the DSO and ISO which is compatible with the current structure of the wholesale market without introducing additional changes to existing wholesale market clearing procedure.
	\item In this coordination framework, the DER aggregators participate in the wholesale market through the coordination of the DSO, which ensures the secure operation of the distribution grid. A parametric programming approach is proposed to construct the bid-in cost function of the DSO (to be submitted to the ISO) and run the DSO-level market clearing procedure, which is built upon the offers collected from the DER aggregators.
	\item A market settlement approach is proposed for the DSO, which coordinates with the wholesale market clearing process and ensures the DSO's non-profit characteristic. It is proved that under the proposed ISO-DSO coordination framework, each DER aggregator will receive identical dispatch signals and payments when they participate in the wholesale market through the coordination of the DSO and when they participate in the wholesale market directly with the ISO overseeing all the transmission-level and distribution-level operating constraints.
	\item The parametric-programming-based ISO-DSO coordination enables complete decoupling between the solution process of the ISO and DSO optimization sub-problems. This avoids iterative ISO-DSO communications within each wholesale market clearing interval and allows the ISO and DSO to exchange the minimal amount of public data only after each entity reaches its optimal solution. The proposed approach requires no exchange of private/confidential ISO or DSO grid model data.
	\item The DSO's market outcomes and market settlement process are investigated through case studies on the modified IEEE 33-node and 240-node distribution test systems.
\end{itemize} 

The rest of the paper is organized as follows. Section \ref{Framework of DSO} presents general idea of the DSO and ISO coordination framework. Section \ref{Formulation} presents the mathematical formulation of ISO-DSO coordination framework. Section \ref{marketsettlement} proposes the market settlement approach for the ISO-DSO coordination framework. Section \ref{CaseStudies} discusses the case studies. Section \ref{Conslusion} presents the concluding remarks.

\section{The ISO-DSO Coordination Framework}\label{Framework of DSO}

\begin{figure}%, height = 1.5in
	\centering
	\includegraphics[width=1\columnwidth]{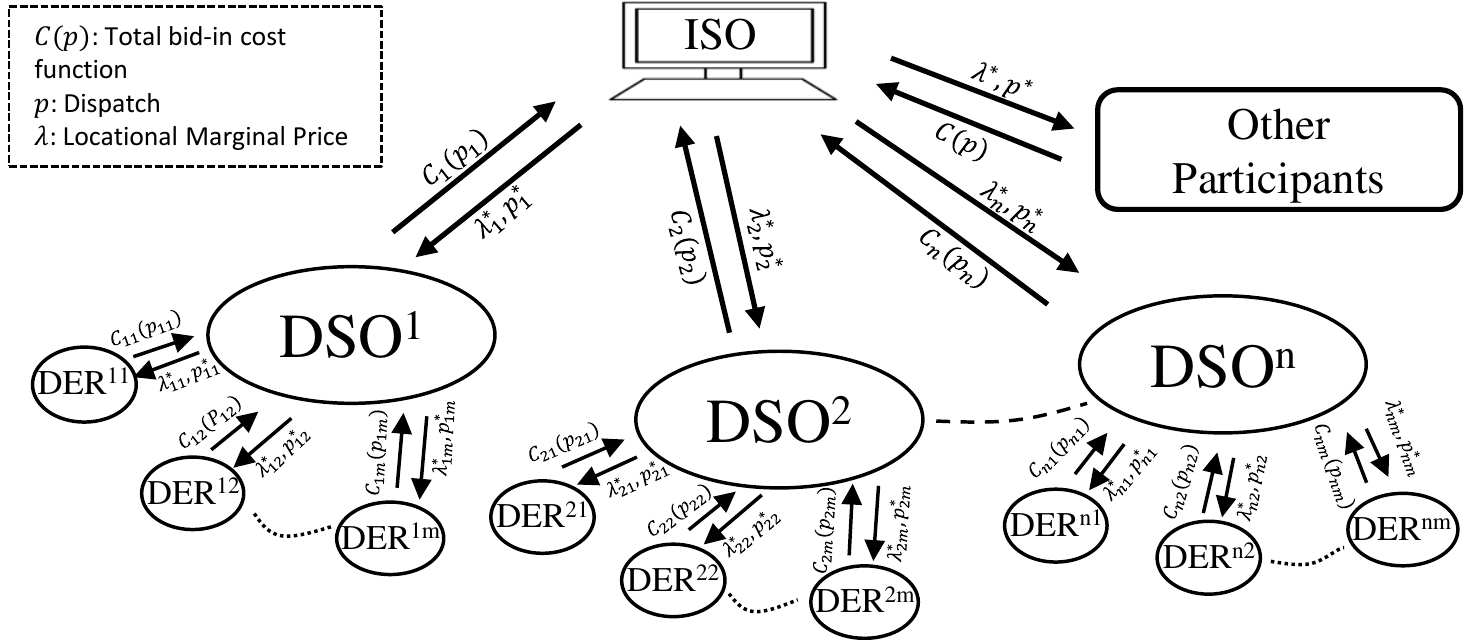}
	\caption{The framework of the ISO-DSO coordination.}\label{fig.11DSO}
\end{figure}

In this section, the proposed ISO-DSO coordination framework, which is shown in Fig. \ref{fig.11DSO}, is explained. In this work, the DSO is defined as a non-profit entity that deals with the wholesale market on one side and coordinates the DER aggregators on the other side. The DER aggregators participate in the wholesale market through the coordination of the DSO, instead of directly participating in the wholesale market. The DER aggregators submit their offers to the DSO. The DSO gathers all these aggregated DER offers and runs the DSO market at the distribution level to construct the bid-in cost function of the DSO using a parametric programming approach. Then, the DSO submits that bid-in cost function to the wholesale market operated by the ISO. The ISO gathers the bid-in cost functions from all the DSOs as well as from other market participants and clears the wholesale market. Then, the ISO sends the dispatches and loactional marginal prices (LMPs) to the DSOs and other market participants. Once each DSO receives the wholesale-level dispatch and LMP from the ISO, the DSO will determine the optimal operating point of the DSO market and determine the optimal dispatches of the DER aggregators in the DSO territory. Then, each DSO will determine the distribution LMPs (D-LMPs) in the distribution system based on the wholesale-level dispatch and LMP received {by the DSO from the ISO at the coupling substation. The DSO will also settle} the DSO market participants (i.e., DER aggregators) based on these D-LMPs. Following this procedure, the optimal dispatches and LMPs for all the ISO-level and DSO-level market participants determined by this ISO-DSO coordination framework will be identical to those determined by the ideal case in which the ISO oversees all the T\&D-level operating constraints.
%This procedure will result in the same optimality conditions as the case in which all the DERs aggregators participate in the wholesale market directly and ISO overseas all the distribution system constraints as well. 
This procedure eliminates the ISO's modeling/computation burden since it avoids sending the distribution grid model data to the ISO by allowing the distribution-level operating constraints and computations to be handled by each DSOs (instead of the ISO).

%of the distribution system for the ISO since the DSO problem can be solved independently before running the wholesale market.

\section{ISO-DSO Coordination Formulation} \label{Formulation}
In this section, the mathematical formulation of the proposed ISO-DSO coordination framework is presented. In order to evaluate the proposed ISO-DSO coordination, an ideal case in which the ISO can oversee all the T\&D-level operating constraints and the DER aggregators can participate in the wholesale market directly is defined and formulated. Then, the formulation of the ISO-DSO coordination is proposed.
\subsection{Ideal Case}
To perform market clearing computations for generating resources in both transmission-level and distribution-level systems (i.e., the conventional generators and DERs), as well as ensuring the secure operation of both T\&D systems, an ideal market framework will be letting one single entity (the ISO) 1) collect both T\&D-level offers/bids (i.e., the bid-in cost functions) from all the conventional generators and DER aggregators; 2) oversee both T\&D-level operating constraints; and 3) optimally dispatch both T\&D-level resources (conventional generators and DER aggregators). 
%In this section, a mathematical formulation of the ideal case is presented. It is defined as a case in which all the DER aggregators can participate in the wholesale market directly. Moreover, the ISO not only considers the transmission level operating constraints but also oversees the \blue{security and physical constraints of all the connected distribution systems}. \blue{This case is considered as the ideal case for the market participation of DER aggregators, since there is one single entity (the ISO) which performs the market clearing computations by modeling and considering all the operating constraints from the transmission system and all the connected distribution systems.}
%Note that, this case is the best case that we can have for the market participation of DER aggregators since every operating constraint from the distribution system up to the transmission system is modeled and considered. 
However, this ideal case is not implementable with the current practice of the wholesale market since 1) the distribution system is not observable to the ISO 2) considering all these small DER aggregators and all the distribution-level constraints in the wholesale market increases the computation burden of the ISO problem. This paper addresses this implementability issue by decomposing this ideal case into one ISO and multiple DSO sub-problems. This decomposition allows the distribution-level modeling and computation burden to be handled by each DSO, such that the ISO only needs to handle transmission-level modeling and computation while coordinating with the DSOs. The following sections prove that the proposed ISO-DSO coordination framework and this ideal case achieve identical optimal dispatches and LMPs for all the T\&D-level market participants (generators and DER aggregators).

%We have defined this ideal case to show that our proposed ISO-DSO coordination framework is as good as the ideal case while removing the challenges of the ideal case.

%In the following, a general formulation of the economic dispatch run by the ISO is presented:
This ideal case is formulated as follows.
\begin{subequations}\label{Formulation1}
\begin{align}
\text{Min}_{\bm{p}} \,\,\,\,\,\,& \sum_{i \in \mathcal{N}_{gen}}c^{g}_{i}(p^{g}_{i})+\sum_{j \in \mathcal{N}_{dis}}\sum_{i \in \mathcal{N}_{agg}}c^{agg}_{i,j}(p^{agg}_{i,j})\label{equ1}\\
\begin{split}\label{equ2}
	\!\!\!\!\!\!\!\!\!\!\!\!	\text{s.t.} \,\,\,\,\,& \bm{p}^{g}\in S^{Tra}\\
& \bm{p}^{agg}_{j}\in S_{j}^{Dis}, \forall j \in \mathcal{N}_{dis}\\
	& p^{g}_{i} \in S_i^{gen}, \forall i \in \mathcal{N}_{gen}\\
	& p^{agg}_{i,j} \in S_{i,j}^{agg}, \forall i \in \mathcal{N}_{agg}, j \in \mathcal{N}_{dis}\\
\end{split}
\end{align}
\end{subequations}
where $i$ and $j$ are indices for market participants (generators/aggregators) and distribution grids in the ISO territory, respectively; $p_{i}^{g}$, $p_{i,j}^{agg}$ and $c^{g}_{i}(p_{i}^{g})$ , $c^{agg}_{i,j}(p^{agg}_{i,j})$ are the dispatched power and bid-in cost functions of generators in the transmission system and aggregators in various distribution systems under the ISO territory, respectively; $\bm{p}$ is the vector of $p_{i}$; $S^{Tra}$, $S^{Dis}_{j}$, $S^{gen}_{i}$, and $S^{agg}_{i,j}$ are the search space defined by the system-wide transmission grid constraints, system-wide distribution grid constraints, operating constraints of generators, and operating constraints of aggregators, respectively; $\mathcal{N}_{gen}$ is the set of all generators; $\mathcal{N}_{dis}$ is the set of all distribution systems; $\mathcal{N}_{agg}$ is set of all aggregators.  

Equation (\ref{equ1}) minimizes the total cost function of the wholesale market considering all the generators and DER aggregators. Equation (\ref{equ2}) presents the operating constraints of all market participants (generators and DER aggregators) as well as the physical constraints of the T\&D systems.

\subsection{ISO-DSO Coordination}
In this section, the mathematical formulation of the proposed ISO-DSO coordination framework is presented. This framework decomposes the above ideal case into one ISO sub-problem and multiple DSO sub-problems. Each DSO sub-problem can be solved independently. This framework and the ideal case will result in identical optimal dispatch and payment/LMP to each of the T\&D-level market participants. However, the decomposition in this framework reduces the computation and modeling burden of the ISO by moving all the distribution-level decision variables and constraints to each DSO's sub-problem.

The ISO sub-problem is formulated as follows:
\begin{subequations}\label{Formulation2}
\begin{align}
	%\begin{split}
\text{Min}_{\bm{p}} \,\,\,\,\,\,&\sum_{i \in \mathcal{N}_{gen}}c^{g}_{i}(p^{g}_{i})+\sum_{j \in \mathcal{N}_{dis}}c^{dso}_{j}( p^{dso}_{j}) \label{equ.3}\\
\begin{split}\label{equ.4}
	\!\!\!\!\!\!\!\!\!\!\!\!	\text{s.t.} \,\,\,\,\,& \bm{p}\in S^{Tra}\\
	& p_{i} \in S_i^{gen}, \forall i \in \mathcal{N}_{gen}\\
	& p_{j} \in S_{j}^{dso}, \forall j \in\mathcal{N}_{dis}\\
\end{split}
%\begin{split}
\end{align}
\end{subequations}
where $p^{dso}_{j}$ is the output power of each DSO; $c^{dso}_{j}( p^{dso}_{j})$ is the bid-in cost function of each DSO; $S_{j}^{dso}$ is the search space defined by the operating constraints of each DSO.

Equation (\ref{equ.3}) minimizes the total cost in the wholesale market, after collecting the bid-in cost functions from all the wholesale market participants (including conventional generators and DSOs). Equation (\ref{equ.4}) models all the operating constraints of the DSOs and other wholesale market participants as well as the physical constraints of the transmission system. This ISO sub-problem is compatible with the current wholesale market clearing practice.

Each DSO $j$ needs to determine its bid-in cost function $c_j^{dso}(p_j^{dso})$ and the corresponding DSO operating constraints $S_{j}^{dso}$
%	$S_j^{Dis}$ 
to be submitted to the wholesale market (the above ISO sub-problem). We propose the following parametric programming approach for each DSO $j$ to determine these data.
	
\begin{align}\label{equ.5}
	\begin{split}
		%\begin{split}
		c_{j}^{dso}(p_{j}^{dso})=\text{Min}_{\bm{p}^{agg}} &\sum_{i \in \mathcal{N}_{agg}}c_{i,j}^{agg}(p_{i,j}^{agg})\\
		\text{s.t.} \,\,\,\,\,\,\,& p_{j}^{dso} = \sum_{i \in \mathcal{N}_{agg_k}}p_{i,j}^{agg}+\sum_{l \in \mathcal{N}_{f_k}}f_{l,j}-L_{k,j}\\
		& p_{i,j}^{agg} \in S_{i,j}^{agg},\forall i \in \mathcal{N}_{agg}\\
		&  \bm{p}_{j}^{agg} \in S^{Dis}_j
	\end{split}
\end{align}
where $k$ is the substation node of DSO $j$; $l$ is the index for branches; $\mathcal{N}_{f_k}$ is the set of all branches connected to node $k$; $f_{l,j}$ is the flow on branch $l$ of DSO $j$;  $L_{k,j}$  is the firm load of node $k$ in DSO $j$. 

Equation (\ref{equ.5}) defines a parametric programming problem, where $p_{j}^{dso}$ is treated as a parameter in the minimization problem. Note that while the power balance equation at the substation is explicitly presented as the equality constraint, the power balance constraints in other nodes are incorporated in $S^{Dis}_j$. After collecting bid-in cost functions from all the DSO-level market participants (i.e., DER aggregators), for every possible $p_{j}^{dso}$ value, this problem minimizes the total generation cost in the DSO-level market while satisfying the following constraints: 1) power balance constraint at each node; 2) the operating constraints of DER aggregators; and 3) the system-wide distribution grid constraints. Linearized three-phase power flow is considered which ignores losses \cite{Gan}. 

Before the ISO market clearing run, each DSO collects the bid-in cost functions from all the DSO-level market participants (i.e., the DER aggregators) in its territory and solves (\ref{equ.5}) to determine its DSO bid-in cost function $c_j^{dso}(p_j^{dso})$ and the corresponding DSO operating constraints $S_{j}^{dso}$
%\blue{$S_j^{Dis}$} 
(i.e., the upper/lower limits for the DSO bid-in cost function) to be submitted to the wholesale market in the ISO sub-problem. The ISO collects the bid-in cost functions from all the DSOs and other ISO-level market participants (such as conventional generators), which allows the ISO to clear the wholesale market by solving (\ref{equ.3})-(\ref{equ.4}).

\begin{lemma}\footnote[1]{Proofs of all lemmas and theorems are removed due to space limitations.}
	The optimal bid-in cost function from DSO to ISO, $c_{j}^{dso}(p_{j}^{dso})$, is a convex function of parameter $p_{j}^{dso}$, if the following conditions are all satisfied: 1) the bid-in cost function submitted by each aggregator $c_{i,j}^{agg}(p_{i,j}^{agg})$ is a convex function; 2) the operating constraints of each DER aggregator define a convex set $S_{i,j}^{agg}$; and 3) the system-wide distribution grid constraints define a convex set $S_j^{Dis}$.
\end{lemma}
%\begin{proof}\renewcommand{\qedsymbol}{}
%	\blue{See Appendix \ref{Appendix01}.} \red{Please write a brief proof in the appendix based on \cite{borrelli2003geometric}}
%\end{proof}

The convexity of the optimal DSO bid-in cost function $c_j^{dso}(p_j^{dso})$ ensures that our proposed ISO-DSO coordination is compatible with the current wholesale market structure, by allowing each DSO to always submit a convex bid-in cost function to the ISO. The ISO can then directly clear the wholesale market following its current market clearing procedure without introducing any additional change.

% Let's open the general economic dispatch problem of the DSOs proposed in Problem (5):   
After the ISO clears the wholesale market, the dispatch and LMP data is distributed to all the DSOs. Each DSO utilizes the LMP of the coupling substation, as determined by the ISO, and employs it as the wholesale market price in the subsequent proceedings.
\begin{align}\label{equ.5_1}
		\begin{split}
			%\begin{split}
		\text{Min}_{\bm{p}^{agg}, p_{j}^{dso}} &\sum_{i \in \mathcal{N}_{agg}}c_{i,j}^{agg}(p_{i,j}^{agg})+LMP^{*}_jp_{j}^{dso}\\
			\text{s.t.} \,\,\,\,\,\,\,&p_{j}^{dso} = \sum_{i \in \mathcal{N}_{agg_k}}p_{i,j}^{agg}+\sum_{l \in \mathcal{N}_{f_k}}f_{l,j}-L_{k,j}\\
			& p_{i,j}^{agg} \in S_{i,j}^{agg},\forall i \in \mathcal{N}_{agg}\\
			&  \bm{p}_{j}^{agg} \in S^{Dis}_j
		\end{split}
\end{align}
where 
%$p_{j}^{dso}$ is a decision variable also in this DSO market clearing problem; 
$LMP_j^*$ is the optimal wholesale LMP determined by the ISO market clearing at the bus where the DSO $j$ is located. 

Section \ref{marketsettlement} presents the detailed DSO market settlement procedure for each DSO to utilize (\ref{equ.5}) and (\ref{equ.5_1}) to determine the optimal dispatch and distribution LMPs (D-LMPs) for all the aggregators in the DSO territory.

%\yellow{By solving} (\ref{equ.5_1}), each DSO will clear its DSO market by determining \purple{the dispatch of all the aggregators} and the distribution D-LMPs within its territory. \blue{In (\ref{equ.5_1}), $p_{j}^{dso}$ is a decision variable.} %Meanwhile, it is also a decision variable in the ISO sub-problem in (\ref{Formulation2}). The relationship between $p_{j}^{dso}$ in the ISO and DSO sub-problems is discussed in Section \ref{marketsettlement}.}

A detailed formulation for the above DSO sub-problem in (\ref{equ.5})-(\ref{equ.5_1}) which considers the real/reactive power flow limits and voltage limits using the linearized three-phase distribution power flow \cite{Gan} is presented in Appendix \ref{Appendix01}.

\section{Market Settlement}\label{marketsettlement} 

In our proposed ISO-DSO coordination framework, the DSO is a nonprofit mediator that deals with the DER aggregators on one hand and trades with the wholesale market on the other hand. The DSO gathers the offers from all the DER aggregators and constructs the DSO bid-in cost function and submits it to the ISO based on the parametric programming procedure in (\ref{equ.5}). Once the ISO receives the bid-in cost functions from all the DSOs and other wholesale market participants, the ISO clears the wholesale market by solving (\ref{Formulation2}) and determines the power dispatch $p_j^{dso*}$ and LMP at the ISO-DSO coupling substation for each DSO. The DSO then needs to clear the DSO-level market with $p_j^{dso*}$ and the wholesale-level LMP it receives at the ISO-DSO coupling substation. Each DSO performs this market settlement procedure by 1) letting $p_j^{dso} = p_j^{dso*}$ in (\ref{equ.5}) and solving (\ref{equ.5}) for the optimal dispatch of all the aggregators in the DSO territory when $p_j^{dso} = p_j^{dso*}$; 2) solving (\ref{equ.5_1}) and obtaining the dual variables of (\ref{equ.5_1}) as the optimal D-LMPs of all the aggregators in the DSO territory.
%	\blue{ as well as solving and determining dual variables of node balance constraints in (\ref{equ.5_1})} . 
%This allows the DSO to fix the parameter $p_j^{dso}$ in the parametric-programming-based DSO-level economic dispatch problem at its optimal value determined by the wholesale market clearing. The optimal DSO-level economic dispatch outcomes at $p_j^{dso} = p_j^{dso*}$ can then be obtained by solving (\ref{equ.5}). 
%These optimal outcomes include the optimal power dispatch \blue{from (\ref{equ.5}) and optimal D-LMPs,  dual variables of (\ref{equ.5_1}) for all the DER aggregators in the DSO territory.}
 We prove the theorems below which guarantees that following the above market settlement procedure, the optimal dispatches, LMPs (or D-LMPs), and payments received by all the ISO-level and DSO-level market participants under the proposed ISO-DSO coordination framework will be identical to those under the ideal case where the ISO serves as the single entity overseeing all the T\&D-level market participants and operating constraints.

\setcounter{theorem}{0}
\begin{theorem}\footnotemark[1]
	The optimal dispatches for all the ISO-level and DSO-level market participants under the ISO-DSO coordination framework in (\ref{Formulation2})-(\ref{equ.5}) are identical to those under the ideal case in (\ref{Formulation1}).
\end{theorem}
%\begin{proof}\renewcommand{\qedsymbol}{}
%	See Appendix \ref{Appendix2}.	
%\end{proof}

\begin{theorem}\footnotemark[1]
%	If a DER aggregator had participated directly in the wholesale market operated by the ISO, would have received the same settlement as participated in the DSO market.
	The optimal payments and LMPs (or D-LMPs) for all the ISO-level and DSO-level market participants under the ISO-DSO coordination framework in (\ref{Formulation2})-(\ref{equ.5_1}) are identical to those under the ideal case in (\ref{Formulation1}).
\end{theorem}

%\begin{proof}\renewcommand{\qedsymbol}{}
%	See Appendix \ref{Appendix}.	
%\end{proof}

The above theorems further guarantee: 1) Our proposed ISO-DSO coordination framework completely decouples the optimization problem in the ideal case into one ISO and multiple DSO sub-problems. 2) After this decoupling, at each market clearing run, each DSO only needs to submit its convex bid-in cost function $c_j^{dso}(p_j^{dso})$ and the corresponding DSO operating constraints (upper/lower limits) of this cost function $S_j^{Dis}$ to the ISO, and the ISO only needs to send the optimal wholesale-level dispatch $p_j^{dso*}$ and wholesale LMP back to each DSO. There is no data exchange between different DSOs and no exchange of confidential ISO or DSO grid models. This data exchange procedure is compatible with the current wholesale market clearing practice. It will result in the minimal amount of ISO-DSO data exchange without changing the existing wholesale market clearing procedure. Besides, this data exchange procedure also allows the ISO and DSO to exchange data only after each entity reaches its optimal solution. There is no iterative ISO-DSO data exchange during the iterative solution process of the ISO and DSO sub-problems. This ensures a complete decouple between the iterative solution process of the ISO and DSO sub-problems and eliminates the need for iterative ISO-DSO communications within each market clearing run.

\section{Case Studies}\label{CaseStudies}
In this section, case studies have been implemented to verify the effectiveness of the proposed ISO-DSO coordination model. First, a small illustrative example is studied to clearly describe our proposed approach. Then, a large system is studied which includes an IEEE 118-bus test system in the wholesale market and two distribution systems - the IEEE 33-node balanced and 240-node unbalanced distribution systems.  
\subsection{Illustrative Example}
\begin{figure}[b]
	\centering
	\includegraphics[width=1\columnwidth]{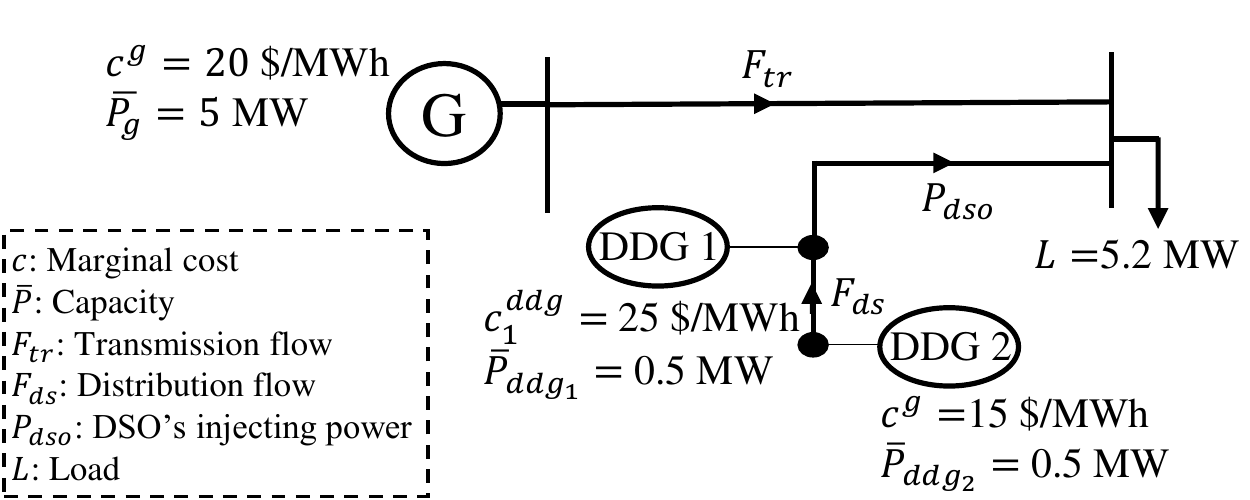}
	\caption{Illustrative example system. The minimum active power for all units is zero. }\label{fig.1.Example}
\end{figure}
\begin{figure}[b]
	\centering
	\includegraphics[width=1\columnwidth]{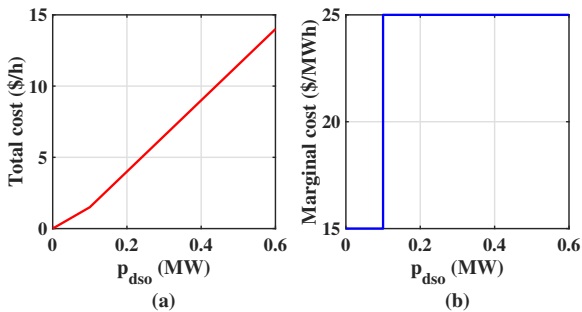}
	\caption{DSO bid-in total (left) and marginal (right) cost functions in the illustrative example.}\label{fig.2.bidcost}
\end{figure}
In this section, in order to understand the proposed ISO-DSO coordination clearly, a small illustrative example is given. The system consists of a generating unit ($G$) and a firm load ($L$) on the transmission side, as well as two dispatchable distributed generations (DDGs) on the distribution side. The system and its corresponding data are provided in Figure \ref{fig.1.Example}. 
The DSO parametric programming problem is as follows:

\begin{subequations}\label{equ.62_1}
\begin{align}
c^{dso}(P_{dso})=\, \text{Min}_{\bm{P}_{ddg}} \,\,\,\, &25P_{ddg_1}+15P_{ddg_2}\\
%  \nonumber\\
\text{s.t.} \,\,\,\,\, & P_{ddg_1}+F_{ds}=P_{dso} \\
& P_{ddg_2}-F_{ds}=0\\
&0\le P_{ddg_1}\le0.5\\
&0\le P_{ddg_2}\le0.5\\
&-0.1\le F_{ds}\le 0.1
\end{align}
\end{subequations}
where $c^{dso}(P_{dso})$ is the bid-in cost function of the DSO; $P_{dso}$ is the output power of the DSO injected; $P_{ddg_1}$ is the active power provided by DDG 1; $P_{ddg_2}$ is the active power provided by DDG 2;  $F_{ds}$ is the distribution line flow.

%\blue{The KKT conditions of the above problem are as follows}:
%
%\begin{subequations}\label{equ.63}
%\begin{align}
%	& 25- \lambda_1+\alpha_1-\alpha_1=0\label{lmp1}\\
%	& 15- \lambda_2+\alpha_2-\alpha_2=0\label{lmp2}\\
%	& P_{dso}-P_{ddg_1}-F_{ds}=0\\
%	& F_{ds}-P_{ddg_2}=0\\
%	& \alpha_1^{+} (P_{ddg_1}- 0.5)=0\label{ddg1-1}\\
%	& \alpha_1^{-} (-P_{ddg_1})=0\label{ddg1-2}\\
%	& \alpha_2^{+} (P_{ddg_2}- 0.5)=0\label{dddg2-1}\\
%	& \alpha_2^{-} (-P_{ddg_2})=0\label{dddg2-2}\\
%	& \mu^{+} (F_{ds}-0.1)=0\\
%	& \mu^{-} (-F_{ds}-0.1)=0
%\end{align}
%\end{subequations}

The problem described above is simple enough that we can determine the bid-in cost function by the following straightforward approach. We simply need to increase the $P_{dso}$ and determine which DDG will provide power and at what cost. As we begin to increase $P_{dso}$, DDG 2 will be the cheaper option, so we can continue to increase $P_{dso}$ until DDG 2 reaches its maximum output or until line $F_{ds}$ becomes congested. Since the capacity of $F_{ds}$ is 0.1 MW, which is lower than the capacity of DDG 2, we can increase $P_{dso}$ up to 0.1 MW, and the cost function would be $c^{dso}(P_{dso})=15P_{dso}$, which is determined by DDG 2. If we need to increase $P_{dso}$ beyond 0.1 MW, we must use DDG 1. We can increase $P_{dso}$ until DDG 1 reaches its maximum output of 0.5 MW. Thus, we can increase $P_{dso}$ up to 0.6 MW, and the total cost function would be $c^{dso}(P_{dso})=15\times0.1+25(P_{dso}-0.1)$. Therefore, the total cost function is as follows:
\[
c^{dso}(P_{dso}) = 
\left\{
\begin{aligned}
	&15P_{dso}, \hspace{-3mm}&& P_{dso}\in [0,0.1)\\
&15\times0.1+25(P_{dso}-0.1), \hspace{-3mm}&& P_{dso}\in [0.1,0.6] \\
\end{aligned}
\right.
\]
% \blue{If $P_{dso} \in [0,0.1)$, DDG 2 will provide the power for the DSO ($P_{ddg_2} \in [0,0.1)$ and $P_{ddg_1}=0$) as it is the cheaper unit. In this case, the cost function is given by $c^{dso}(p^{dso})=15P_{dso}$. Since $P_{ddg_2}$ has a lower limit of 0 and an upper limit of 0.1, we have $\alpha^{-}_2=\alpha^{+}_2=0$ as per (\ref{dddg2-1}) and (\ref{dddg2-2}). Also, from (\ref{ddg1-1}), we get $\alpha^{+}_1=0$. Substituting these values in (\ref{lmp1}) and (\ref{lmp2}), we obtain $\lambda_1=\lambda_2=15$ \$/MWh. }
%
%\blue{When $P_{dso} \in (0.1,0.6)$, the distribution flow $F_{ds}$ is at its maximum and DDG 2 is unable to provide any additional power. DDG 1 will then provide the remaining power required by the DSO ($P_{ddg_2}=0.1$ and $P_{ddg_1}\in [0,0.5)$). In this case, the cost function is given by $c^{dso}(p^{dso})=15\times0.1+25(P_{dso}-0.1)$. From (\ref{dddg2-1}) and (\ref{dddg2-2}), it follows that $\alpha^{-}_2=\alpha^{+}_2=0$. From (\ref{ddg1-2}), we have $\alpha^{-}_1=0$. Consequently, using (\ref{lmp1}) and (\ref{lmp2}), we obtain $\lambda_1=25 $\$/MWh and $\lambda_2=15 $\$/MWh.}

Hence, the bid-in total cost function and marginal cost function which is derivative of the total cost function are determined as shown in Fig. \ref{fig.2.bidcost}(a) and Fig. \ref{fig.2.bidcost}(b), respectively.

The DSO submits this marginal cost function in Fig. \ref{fig.2.bidcost}(b) to the wholesale market and then, the wholesale market runs the following ISO-level economic dispatch problem:

\begin{subequations}\label{Exp2}
\begin{align}\label{equ.64}
	\text{Min}_{\bm{P}}& \,\,\,\,20P_{g}+15P_{dso,1}+25P_{dso,2}\\
	 s.t. &\,\,\,\, P_{g}-F_{tr}=0 \hspace{2.85cm} [\lambda^{WM}_1]\\
	&\,\,\, F_{tr}+P_{dso,1}+P_{dso,2}=5.2\hspace{0.8cm} [\lambda^{WM}_2]\\
	&\,\,\, 0\le P_{g}\le5\hspace{2.5cm} \\
	&\,\,\, 0\le P_{dso,1}\le0.1\hspace{2.5cm} \\
	&\,\,\, 0\le P_{dso,2}\le 0.5\hspace{2.1cm} 
\end{align}
\end{subequations}
where $P_{g}$ is the power provision from the transmission side unit; $P_{dso,1}$ and $P_{dso,2}$ are the power provision of the first and second segments of the DSO bid-in cost function shown in Fig. \ref{fig.2.bidcost}(b), respectively; $F_{tr}$ is the transmission line flow; $\lambda^{WM}_1$ and $\lambda^{WM}_2$ are the dual variables corresponding to the transmission-level power balance constraints, respectively.

The optimal solution to the above ISO problem is: $P_{g}=5$ MW, $P_{dso,1}=0.1$ MW, $P_{dso,2}=0.1$ MW, and 
the DSO has dispatched $0.1+0.1= 0.2$ MW and the wholesale LMP at the ISO-DSO coupling bus is $25$ \$/MWh. The DSO substitutes the parameter $P_{dso}=0.2$ in (\ref{equ.62_1}) and determines the DDGs' optimal dispatches, $P_{ddg_1}=0.1$ MW, $P_{ddg_2}=0.1$ MW. Then, the DSO solves the following problem to determine optimal D-LMPs: 
\begin{subequations}
	\begin{align}\label{equ.62}
		\text{Min}_{\bm{P}} \,\,\,\, &25P_{ddg_1}+15P_{ddg_2}+25P_{dso}\\
%		&\text{s.t.} \nonumber\\
		\text{s.t.} \,\,\,\, & P_{ddg_1}+F_{ds}=P_{dso} \hspace{2cm} [\lambda_1]\\
		& P_{ddg_2}-F_{ds}=0\hspace{2.5cm} [\lambda_2]\\
		&0\le P_{ddg_1}\le0.5\hspace{2.5cm} \\
		&0\le P_{ddg_2}\le0.5\hspace{2.5cm} \\
		&-0.1\le F_{ds}\le 0.1\hspace{2.1cm} 
	\end{align}
\end{subequations}
where $\lambda_{1}, \lambda_2$ are the dual variables corresponding to the distribution-level power balance constraints (i.e., the D-LMPs).
%If the wholesale market were to solve the problem considering the direct participation of DER aggregators, its problem is as follows:\\
The solution to this DSO problem determines the following D-LMPs: $\lambda_1=25$ \$/MWh, and $\lambda_2=15$ \$/MWh. 

The following equations describe the ideal case in which the ISO can oversee both T\&D-level operations and DER aggregators directly participate in the ISO market:

\begin{subequations}\label{Exp3}
\begin{align}\label{equ.65}
	\text{Min}_{\bm{P}} \,\,\,\, & 20P_{g}+15P_{ddg1}+25P_{ddg2}\\
	 s.t. &\,\,\,\, P_{g}-F_{tr}=0  &[\lambda^{WM}_1]\\
	&\,\,\, F_{tr}+P_{dso}=5.2 &[\lambda^{WM}_2]\\
	&\,\,\, P_{ddg_1}+F_{ds}=P_{dso} &[\lambda_1]\label{lambda1}\\
	&\,\,\, P_{ddg_2}-F_{ds}=0 &[\lambda_2]\label{lambda2}\\
	&\,\,\, 0\le P_{g}\le5\hspace{2.3cm} \\
	&\,\,\, 0\le P_{ddg1}\le0.5\hspace{2.5cm} \\
	&\,\,\, 0\le P_{ddg2}\le 0.5\hspace{2.1cm} \\
	&\,\,\, -6\le F_{tr}\le 6\\
	&\,\,\, -0.1\le F_{ds}\le 0.1\label{ds_flow}
\end{align}
\end{subequations}

The solution (including optimal dispatch and prices for all the T\&D-level resources) to the above ideal case is the same as the ISO-DSO coordination framework. However, upon comparing formulations (\ref{Exp2}) and (\ref{Exp3}), it can be observed that constraints (\ref{lambda1}), (\ref{lambda2}), and (\ref{ds_flow}) are no longer necessary, which reduces the problem size and computational burden for the ISO, as well as avoids sending DSO-level modeling details to the ISO.
	
%	resulting in a reduction in problem size and computational burden for the ISO.}

\subsection{Large Test System}
In this section, simulation studies are implemented in a large test system containing ISO running the wholesale-level economic dispatch on an IEEE 118-bus test system. We have also considered two DSOs running the DSO-level market in the IEEE 33-node balanced and 240-node unbalanced distribution systems, respectively. YALMIP \cite{Lofberg2004} is utilized to solve parametric programming problems for DSOs. 
\subsubsection{118-bus Test System Data}
The IEEE 118-bus test system is considered as the transmission system operated by the ISO. The system data is given in \cite{IEEE118}. The system contains 118 buses, 186 transmission lines, and 54 generators.  
\subsubsection{33-node Test System Data and Results}
\begin{figure}%, height = 1.5in
	\centering
	\includegraphics[width=1\columnwidth]{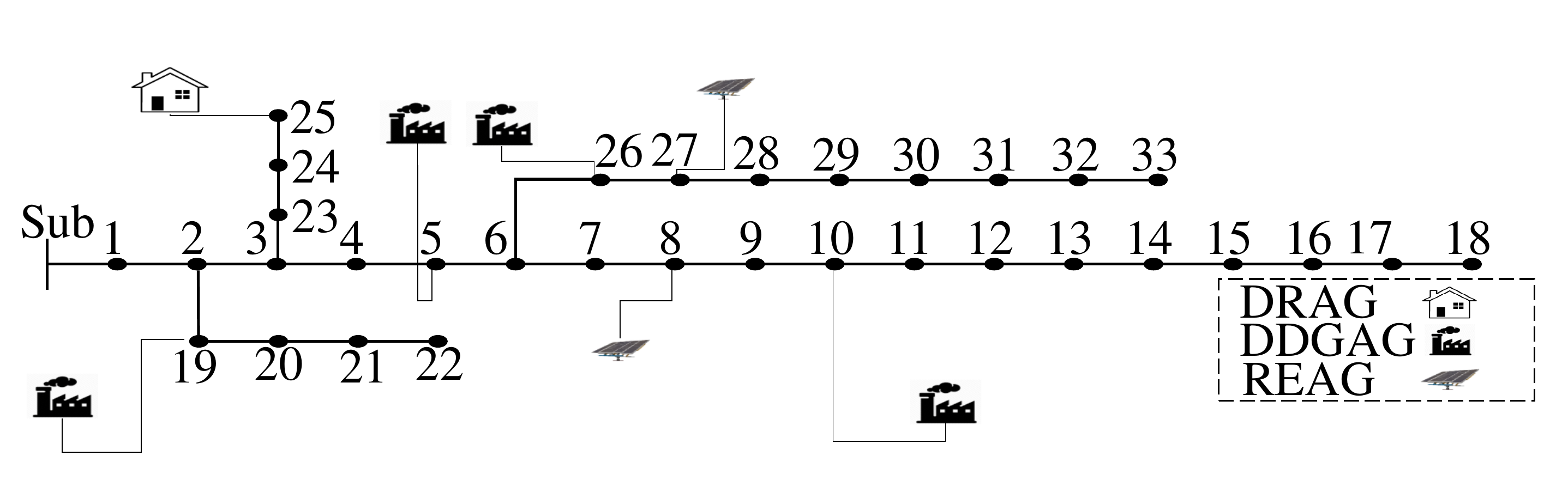}
	\caption{33-node test system.}\label{fig.33node}
\end{figure}
\begin{table}
	\centering
	\caption{DSO market participants data for the 33-node test system }\label{table.1 input data}
	\begin{tabular}{c|c|c|c}
		\hline
		
		Participant&Pmin (MW)&Pmax (MW) &Offering price (\$/MWh)\\
		\hline
		\hline
		DDGAG 1&0&0.5 &20\\
		\hline
		DDGAG 2&0&1 &10\\
		\hline
		DDGAG 3&0&1.2 &15\\
		\hline	
		DDGAG 4&0&2 &24\\
		\hline		
		DRAG &0&2 &28\\
		\hline
	\end{tabular}
\end{table}
\begin{figure}%, height = 1.5in
	\centering
	\includegraphics[width=1\columnwidth]{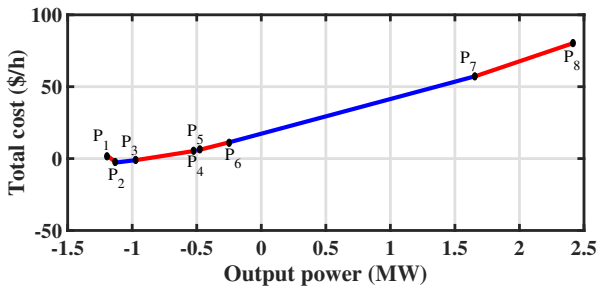}
	\caption{Total cost function of the 33 node test system.}\label{fig.totalcost_33}
\end{figure}
\begin{figure}%, height = 1.5in
	\centering
	\includegraphics[width=1\columnwidth]{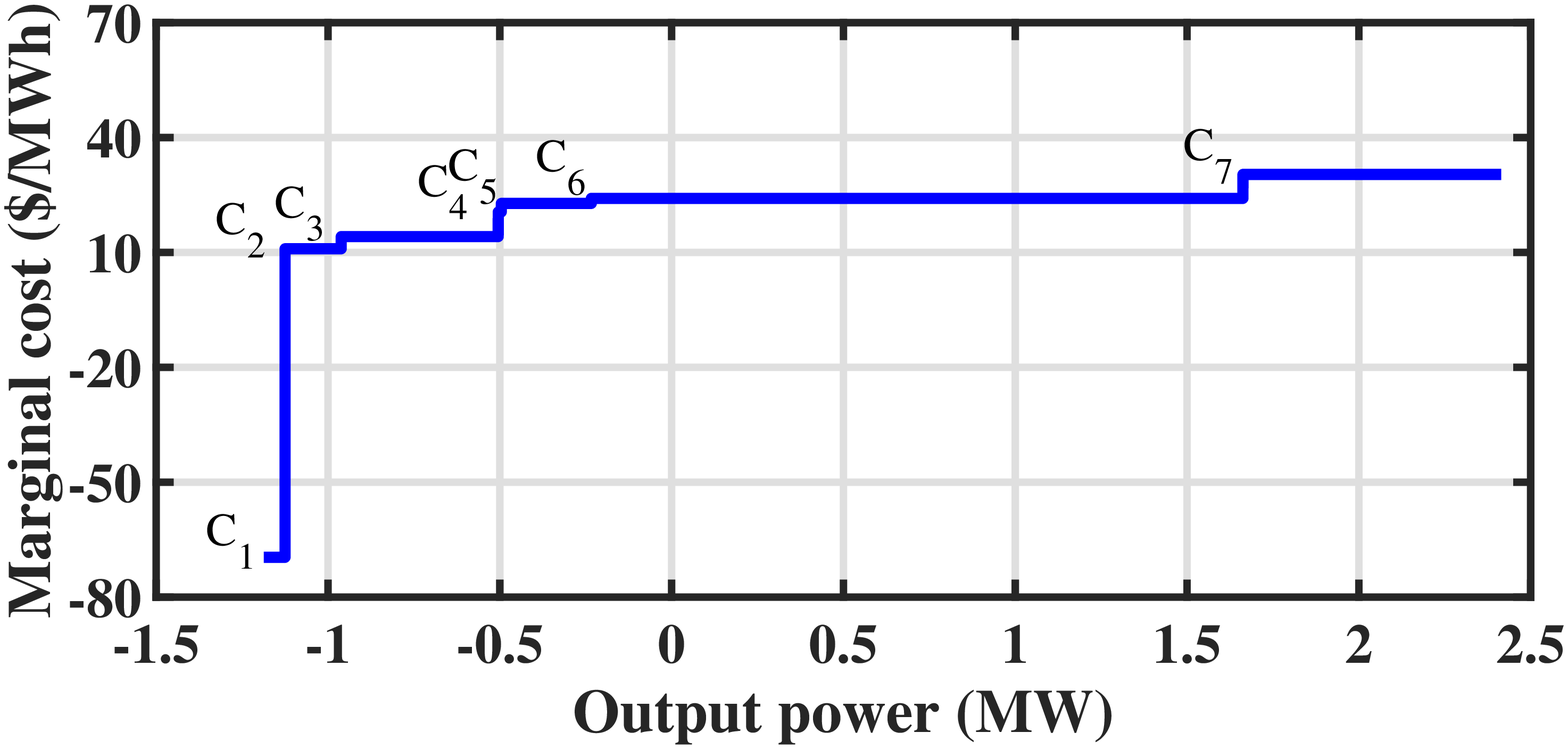}
	\caption{Bid-in marginal cost function of the 33 node test system.}\label{fig.bidincost_33}
\end{figure}

\begin{table}
	\centering
	\caption{33-node test system breakpoints and marginal costs data}\label{table.2breakpoints}
	\begin{tabular}{c|c|c|c}
		\hline
		\multirow{2}{1.2cm}{Breakpoint index}&\multirow{2}{2.6 cm}{Breakpoint coordinate value (MW,\$/h)}&\multirow{2}{1.2 cm}{Marginal cost index}&\multirow{2}{1.8 cm}{Marginal cost value (\$/MWh)}\\ 
		&&&\\
		\hline
		\hline
		P$_1$&(-1.18654,1.54166)&C$_1$ &-69.6072\\
		\hline
		P$_2$&(-1.12498, -2.74336)&C$_2$&10.9555\\
		\hline
		P$_3$&(-0.961451, -0.951813)&C$_3$ &14.1388\\
		\hline	
		P$_4$&(-0.504623, 5.50717)&C$_4$ &20.5587\\
		\hline		
		P$_5$ &(-0.495727, 5.69006)&C$_5$ &22.8007\\
		\hline
		P$_6$ &(-0.233645, 11.6657)&C$_6$ &24.1164\\
		\hline
		P$_7$ &(1.66269, 57.3985)&C$_7$ &30.3739\\
		\hline
		P$_8$ &(2.4175, 80.325)& &\\
		\hline
	\end{tabular}
\end{table}
The 33-node test system is a balanced radial network which is shown in Fig. \ref{fig.33node}. The system contains 33 nodes, 32 branches, a demand response aggregator (DRAG),  four dispatchable distributed generation aggregators (DDGAGs), and two renewable energy aggregators (REAGs). The test system data and load data are given in \cite{MM}. The two REAGs are considered to have identical energy production profiles of 1 MW. The other aggregators' data is given in Table. \ref{table.1 input data}. Pmin and Pmax are the minimum and maximum generating power, respectively. It is assumed that the 33-node test system is connected to the 118-bus test system trough bus 87 on the transmission side.

The total cost function of the DSO is determined based on (\ref{Formulation4}). The DSO's total (minimal) operating costs at different output power levels are shown in Fig. \ref{fig.totalcost_33}. This is a piecewise linear function with eight breakpoints separating the seven linear segments.
%Considering the minimum and maximum points as breakpoints, in total, seven breakpoints are starting from P$_1$ and ending at P$_7$ in Fig. \ref{fig.totalcost_33}. 
The breakpoints in Fig. \ref{fig.totalcost_33} are determined by the DSO-level market participants' minimum and maximum output power considering the network's physical constraints. 
The bid-in marginal cost function which is the derivative of the total bid-in cost function in Fig. \ref{fig.totalcost_33} is shown in Fig. \ref{fig.bidincost_33}. This marginal cost function consists of seven levels of marginal costs corresponding to the seven linear segments in the piecewise linear total cost function in Fig. \ref{fig.totalcost_33}.

The coordinates of the breakpoints in Fig. \ref{fig.totalcost_33} and the values of the marginal costs in Fig. \ref{fig.bidincost_33} are given in Table \ref{table.2breakpoints}.

The bid-in marginal cost function starts with the output power of -1.18654 MW which means that DSO can consume the energy of -1.18654 MW due to the capability of the DRAG and inelastic load to consume power in the distribution system. The bid-in price of this consumption is -69.6072 \$/MWh. The negative value indicates that if the wholesale market dispatches this consumption value to the DSO, the DSO should be paid at this price. 
%	In other words, consuming energy at this value will result in costs for the DSO instead of providing benefits.
This indicates the DSO prefer not purchasing energy from the ISO at this segment, since this may increase the total DSO-level generation cost. 
	This is because it may violate certain voltage constraints that require the DSO to provide energy from its costly units. When the price of the wholesale market increases, the DSO starts selling energy to the wholesale market because the price in the wholesale market is higher than the offering prices of the DDGAGs in the distribution system. In the end, the DSO sells energy to the wholesale market at the price of 30.3739 \$/MWh. This is due to the fact that if the offering price of the wholesale market is greater than 30.3739 \$/MWh, the DSO sells the energy to the ISO instead of to the DRAG. The DSO submits its marginal cost function, shown in Fig. \ref{fig.bidincost_33}, to the ISO and waits for the ISO to clear the wholesale market.

\subsubsection{240-node Distribution System Data and Results}
\begin{table}
	\centering
	\caption{DSO market participants information for 240-node test system}\label{table.240nodedata}
	\begin{tabular}{c|c|c|c|c|c}
		\hline
		
		\multirow{2}{0.9 cm}{Participant}&\multirow{2}{0.9 cm}{Capacity\\ (MW)}&\multirow{2}{0.9 cm}{Price\\ (\$/MWh)} &\multirow{2}{1 cm}{Participant}&\multirow{2}{0.9 cm}{Capacity\\ (MW)}& \multirow{2}{0.9 cm}{Price\\ (\$/MWh)}\\
		&&&&&\\
		\hline
		\hline
		DDGAG 1&0.25 A&20&DRAG 1&0.15 A&28\\
		\hline
		DDGAG 2&0.25 A&10&DRAG 2&0.15 A&29\\
		\hline
		DDGAG 3&0.25 B&15&DRAG 3&0.15 B&30\\
		\hline	
		DDGAG 4&0.25 B &24&DRAG 4&0.15 B&27\\
		\hline		
		DDGAG 5 &0.25 C &14&DRAG 5&0.15 C&26\\
		\hline
		DDGAG 6&0.25 C&15&DRAG 6&0.15 C&25\\
		\hline
		DDGAG 7&0.25 A&16&DRAG 7&0.15 A&24\\
		\hline
		DDGAG 8&0.25 B&17&DRAG 8&0.15 B&22\\
		\hline	
		DDGAG 9&0.25 C&18&DRAG 9&0.15 C&22\\
		\hline		
		DDGAG 10 &0.25 A &19&DRAG 10&0.15 A&23\\
		\hline
	\end{tabular}
\end{table}
\begin{figure}%, height = 1.5in
	\centering
	\includegraphics[width=1\columnwidth]{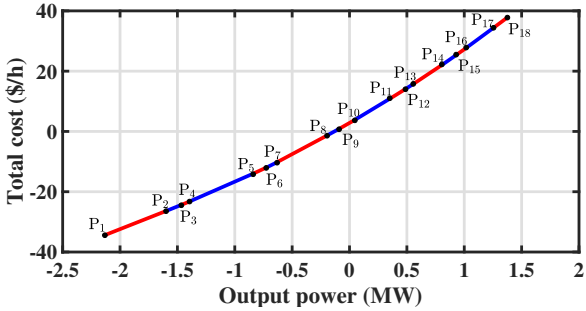}
	\caption{Total cost function of the 240 node test system.}\label{fig.totalcost_240}
\end{figure}

\begin{figure}%, height = 1.5in
	\centering
	\includegraphics[width=1\columnwidth]{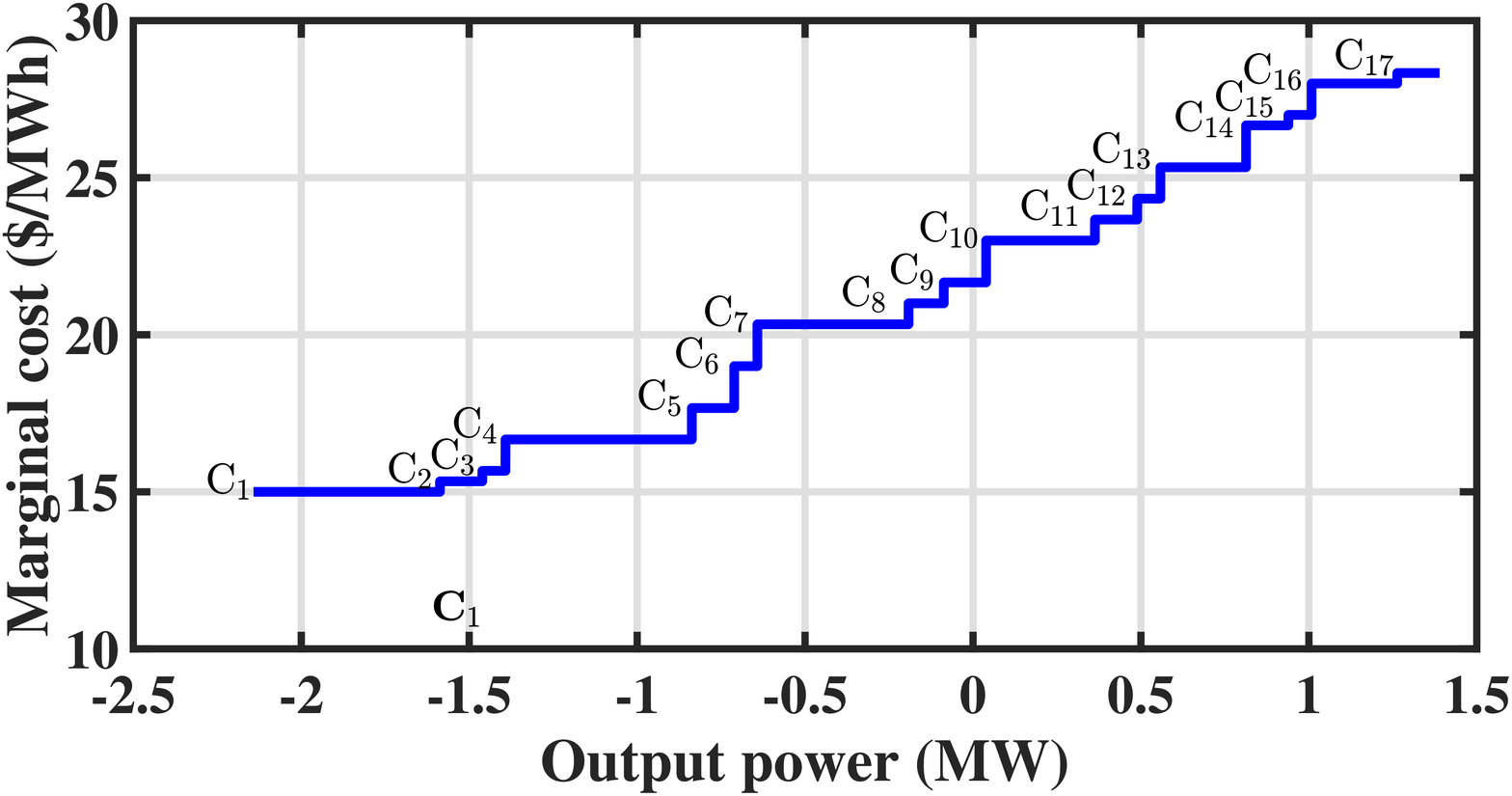}
	\caption{Bid in marginal cost function of the 240 node test system.}\label{fig.bidincost_240}
\end{figure}
\begin{table}
	\centering
	\caption{240 node breakpoints and marginal costs data}\label{table.4breakpoints}
	\begin{tabular}{c|c|c|c}
		\hline
		\multirow{2}{1.2cm}{Breakpoint index}&\multirow{2}{2.6 cm}{Breakpoint coordinate value (MW,\$/h)}&\multirow{2}{1.2 cm}{Marginal cost index}&\multirow{2}{1.8 cm}{Marginal cost value (\$/MWh)}\\ 
		&&&\\
		\hline
		\hline
		P$_1$&(-2.142,-34.538)&C$_1$ &15\\
		\hline
		P$_2$&(-1.587, -26.213)&C$_2$&15.333\\
		\hline
		P$_3$&(-1.461, -24.281)&C$_3$ &15.667\\
		\hline	
		P$_4$&(-1.392, -23.2)&C$_4$ &16.667 \\
		\hline		
		P$_5$ &(-0.837, -13.95)&C$_5$ &17.667\\
		\hline
		P$_6$ &(-0.711, -11.724)&C$_6$ &19\\
		\hline
		P$_7$ &(-0.642, -10.413)&C$_7$ &20.333\\
		\hline
		P$_8$&(-0.192,-1.263)&C$_{8}$ &21\\
		\hline
		P$_9$&(-0.087, 0.942)&C$_{9}$&21.667\\
		\hline
		P$_{10}$&(0.039, 3.672)&C$_{10}$ &23\\
		\hline	
		P$_{11}$&(0.363, 11.124)&C$_{11}$ &23.667\\
		\hline		
		P$_{12}$&(0.489, 14.106) &C$_{12}$ &24.333\\
		\hline
		P$_{13}$ &(0.558, 15.785)&C$_{13}$ &25.333\\
		\hline
		P$_{14}$ &(0.813, 22.245)&C$_{14}$ &26.667\\
		\hline
		P$_{15}$&(.939,25.605)&C$_{15}$ &27\\
		\hline
		P$_{16}$&(1.008, 27.468)&C$_{16}$&28\\
		\hline
		P$_{17}$&(1.263, 34.608)&C$_{17}$ &28.333\\
		\hline	
		P$_{18}$&(1.389, 38.178)& &\\
		\hline		
	\end{tabular}
\end{table}
The 240-node distribution test system is an unbalanced radial network in Midwest U.S. The data of the system is given in \cite{bu2019time}. 
The system contains 240 nodes and 239 branches. Multiple aggregators are considered as follows: ten DRAGs, ten DDGAGs, and four REAGs. The data of the DER aggregators are given in Table \ref{table.240nodedata}. It is assumed that the 240-node system is connected to the 118-bus system through bus 27 of the transmission system. 

The bid-in cost function of the DSO is determined based on (\ref{Formulation4}). The formulation is extended to handle the single-phase aggregators and unbalanced distribution system physical constraints based on our prior work in \cite{mousavi2020dso}. 

The DSO's total bid-in cost function of the 240-node test system is shown in Fig. \ref{fig.totalcost_240}. The breakpoints in Fig. \ref{fig.totalcost_240} are determined by the DSO market participants' minimum and maximum output power as well as the physical constraints of the distribution system. There are 18 breakpoints including the beginning and ending points. The bid-in marginal cost function of the DSO which is the derivative of the total bid-in cost function in Fig. \ref{fig.totalcost_240} is given in Fig. \ref{fig.bidincost_240}. The data of the breakpoints and the marginal costs are given in Table \ref{table.4breakpoints}.

In Fig. \ref{fig.bidincost_240}, the bid-in marginal cost function starts with -2.14 MW with the price of 15 \$/MWh which means that if the price of the wholesale market is lower than or equal to this value the DSO operating the 240-node test system buys energy from the wholesale market for consumption in the distribution system. As the wholesale market price increases, the energy consumption in the DSO decreases until it reaches 23 \$/MWh at which the DSO sells energy to the wholesale market for any price greater than this value. The amount of energy provision of the DSO for the ISO increases as the price in the wholesale market increases until it reaches its maximum capacity which is 1.39 MW.

\subsubsection{Market Clearing Results}
\begin{table}
	\centering
	\caption{Ideal case and ISO-DSO coordination case dispatch}\label{table5Idealcase}
	\begin{tabular}{c|c|c|c}
		\hline
		\multicolumn{4}{c}{Total wholesale market generators' dispatch }\\
		\hline
		\multicolumn{4}{c}{6601.1 MW}\\
		\hline
		\hline
		\multicolumn{4}{c}{33 node test system dispatches}\\
		\hline
		Participant&Dispatch (MW)&Participant&Dispatch (MW)\\
		\hline
		DDGAG 1&0&DDGAG 3 &1.2\\
		\hline
		DDGAG 2&0.7102&DDGAG 4&0\\
		\hline		
		DRAG &0.6998& &\\
		\hline
		\hline
		\multicolumn{4}{c}{240 node test system dispatches}\\
		\hline
		Participant&Dispatch (MW)&Participant&Dispatch (MW)\\
		\hline
		DDGAG 1&0.065 A&DRAG 1&0.15 A\\
		\hline
		DDGAG 2&0.25 A&DRAG 2&0.15 A \\
		\hline
		DDGAG 3&0.25 B&DRAG 3&0.15 B\\
		\hline	
		DDGAG 4&0 B &DRAG 4&0.15 B\\
		\hline		
		DDGAG 5 &0.25 C &DRAG 5& 0.15 C\\
		\hline
		DDGAG 6&0.25 C&DRAG 6& 0.15 C\\
		\hline
		DDGAG 7&0.25 A&DRAG 7& 0.15 A\\
		\hline
		DDGAG 8&0.25 A&DRAG 8&0.15 A\\
		\hline	
		DDGAG 9&0.25 B&DRAG 9&0.15 B\\
		\hline		
		DDGAG 10 &0.023 C &DRAG 10&0.15 C\\
		\hline
	\end{tabular}
\end{table}
This section compares the market clearing results of the ideal case in (\ref{Formulation1}) and our proposed ISO-DSO coordination case. In the ideal case, the ISO is the single entity which oversees all the market participants and operating constraints in the transmission system and in both distribution systems. In our proposed ISO-DSO coordination case, both 33-node and 240-node DSOs submit their marginal bid-in cost functions in Figs. \ref{fig.bidincost_33} and \ref{fig.bidincost_240} to the ISO. Then, ISO clears the wholesale market based on (\ref{Formulation2}).
Table \ref{table5Idealcase} shows the market dispatch results of the ideal case and the ISO-DSO coordination case for this large test system. Since these two cases share identical market dispatch results for all the T\&D-level market participants (generators and DER aggregators), we only used one table to present these identical results for both cases.
Table \ref{table5Idealcase} shows the total dispatch in the wholesale market and the individual DER aggregators' dispatches in both 33-node and 240-node distribution systems. In the ISO-DSO coordination case, the total dispatches for the 33-node and 240-node DSOs are -0.5046 MW and -0.642 MW, respectively. %\purple{These total DSO dispatches can be obtained by adding the individual dispatches for all the DER aggregators in the same DSO territory. Comment: Does this total dispatch include the firm load in each DSO?}   
\subsubsection{Market Settlements}
In this section, we compare the market settlements of the ideal case and ISO-DSO coordination case. In the ideal case, the LMP on the transmission side is 20.24 \$/MWh, which remains the same throughout the transmission system, as there is no transmission-level congestion. Therefore, the LMPs at the coupling points of the 33 node system and 240-node system are also 20.24 \$/MWh. The 240-node system is unbalanced, resulting in different D-LMPs for each phase, namely 20 \$/MWh, 21.71 \$/MWh, and 19 \$/MWh for phase A, phase B, and phase C, respectively. The average of the three-phase D-LMPs is 20.24 \$/MWh. 
More detailed information on determining LMP in an unbalanced system can be found in our previous work \cite{mousavi2021dso}.

In the ISO-DSO coordination case, the LMP on the transmission side is obtained by (\ref{Formulation2}) and remains identical to the ideal case (20.24 \$/MWh). Each DSO then 
% adds new generation with a generating cost equal to 20.24 \$/MWh and 
determines its own D-LMPs based on (\ref{equ.5_1}), by letting $LMP^*_j = 20.24$ \$/MWh. The D-LMPs of the 33 node test system and 240-node system obtained from the ISO-DSO coordination case are identical to those obtained from the ideal case.
\section{Conclusion}\label{Conslusion} 
In this paper, an ISO-DSO coordination framework is proposed based on parametric programming, which ensures distribution grid operating security while allowing wholesale market participation of DER aggregators.
%novel coordination framework for the operation of the ISO and DSOs is proposed based on parametric programming. 
Each DSO runs the DSO-level market in the distribution system and gathers offers from all the market participants (DER aggregators) in its territory and build the bid-in cost function for submission to the ISO. Then, the ISO gathers all these bid-in cost functions from all the DSOs and from other wholesale market participants to clear the wholesale market.  Once the ISO clears the wholesale market, the dispatch and payment of each DSO are determined. Then, DSOs determine the DSO-level dispatch and D-LMPs in their territories based on the ISO-cleared market. A market settlement approach is presented and proved that each market participant (generator or aggregator) will receive identical compensation and dispatch under the proposed ISO-DSO coordination framework and under the ideal case where the DER aggregators can participate in the wholesale market directly and the ISO is the single entity overseeing both T\&D-level operating constraints. This ISO-DSO coordination framework is compatible with today's wholesale market structure without introducing additional changes to existing wholesale market clearing procedure. It only exchanges minimal amount of public data between the ISO and DSO without exchanging any confidential grid models between the T\&D operations. It also completely decouples the solution process of the ISO and DSO optimization sub-problems, which allows the ISO and DSO to exchange data only after each entity converges to its optimal solution.
%Mathematically proved that the results of the proposed ISO-DSO coordination are the same as the ideal case. 

Case studies were implemented on a small illustrative example and a large system to investigate the proposed ISO-DSO coordination framework. The small illustrative example shows that, compared to the ideal case, the proposed model significantly removes the variables and constraints for the wholesale market while resulting in the same market clearing outcomes. The large system contains the IEEE 118-bus transmission system connected to two DSO operated distribution systems including the 33-node balanced 240-node unbalanced distribution systems. The bid-in cost functions of the DSOs were developed based on parametric programming and submitted to the ISO. The dispatches and payments to the DER aggregators are identical under the ISO-DSO coordination framework and under the ideal case.
%It was shown that the computation burden of the ISO is reduced significantly while having the same accuracy as the ideal case.  
\appendices
\section{}\label{Appendix01}
Following is the detailed formulation of (\ref{equ.5}).
\begin{subequations}\label{Formulation4}
	\begin{align}\label{equ.6}
		%\begin{split}
		c^{dso}(P^{dso})= Min & \sum_{g \in G}\sum_{b\in B}P_{g,b}\pi_{g,b}-\sum_{d \in D}\sum_{b\in B}P_{d,b}\pi_{d,b} 
		%\end{split}
	\end{align}
	\vspace{-18pt}
	\begin{align*}
		\hfilneg \text{s.t.} \hspace{9000pt minus 1fil}
	\end{align*}
	\vspace{-18pt}
	\begin{align}
		%	\text{s.t.}\\
		\begin{split}\label{equ.7}
			& \sum_{d\in D}\sum_{b\in B}H_{n,d}P_{d,b}+H^{sub}_nP^{dso}+L^P_n\\
			&-\sum_{g\in G}\sum_{b\in B}H_{n,g} P_{g,b}+\sum_{j\in J}Pl_{j} A_{j,n} =0;\hspace{3mm} \forall n \in N \\
		\end{split}\\
		\begin{split}\label{equ.8}
			&\sum_{d\in D}\sum_{b\in B}H_{n,d}P_{d,b} tan\phi_{d}+H_{n}^{sub}Q^{dso}+L_{n}^{Q}\\
			& -\sum_{g\in G}\sum_{b\in B}H_{n,g} P_{g,b}tan\phi_{g}+\sum_{j\in J}Ql_{j} A_{j,n} =0 ;\forall n \in N 
		\end{split}\\
		& 0 \le P_{g,b} \le P_{b,g}^{max};\hspace{3mm}\forall b \in B, \forall g \in G\label{equ.9}\\
		& 0 \le P_{d,b} \le P_{d,g}^{max};\hspace{3mm}\forall b \in B, \forall d \in D\label{equ.10}\\
		\begin{split}\label{equ.11}
			& U_{m}=U_{n}-2(r_{j} Pl_{j}+x_{j} Ql_{j} );\hspace{3mm}\forall m\in N,\\
			&\hspace{10mm}\forall n \in N,\, C(m,n)=1 ,\, A(j,n)=1 \\
		\end{split}\\
		& \underline{U} \le U_{n} \le \overline{U} ;\hspace{3mm}\forall n \in N \label{equ.12}\\
		& -Pl^{max} \le Pl_{j} \le Pl^{max};\hspace{3mm} \forall j\in J  \label{equ.13}\\
		& -Ql^{max} \le Ql_{j} \le Ql^{max};\hspace{3mm} \forall j\in J  \label{equ.14}
	\end{align}
\end{subequations}
where $g$ and $G$ represent the index and set of all generating aggregators; $d$ and $D$ represent the index and set of all demand response aggregators; $b$ and $B$ represent the index and set of all production/demand blocks; $j$ and $J$ represent the index and set of all lines; $n$ and $N$ represent the index and set of all nodes; $P^{dso}$ represents the DSO's aggregated offers to the ISO market; $P_{g,b}$ and $P_{d,b}$ represent the energy offers submitted by the generating aggregators and demand response aggregators, respectively, with corresponding prices $\pi_{g,b}$ and $\pi_{d,b}$; $H_{n,d}$, $H_{n,g}$, and $H_{n}^{sub}$ represent the mapping matrices of generating aggregators, demand response aggregators, and substations to node $n$, respectively; $Pl_j$ and $Ql_j$ represent the active and reactive power of branch $j$, respectively; $A_{j,n}$ represents the incidence matrix of branches and nodes; $\phi_g$ and $\phi_d$ represent the phase angle of the generating aggregators and demand response aggregators, respectively; $Q_n^D$ represents the reactive power of the firm load at each node; $L^P_n$ and $L^Q_n$ represent the active and reactive power load at each node; $P^{max}_{g,b}$ and $P^{max}_{d,b}$ represent the maximum production/consumption at each block of the generating aggregators and demand response aggregators, respectively; $U$ represents the square of the voltage of each node; $\underline{U}$ and $\overline{U}$ represent the square of the minimum and maximum permitted voltage values, respectively; $r_j$ and $x_j$ represent the resistance and reactance of the branches; $Pl^{max}$ and $Ql^{max}$ represent the maximum active and reactive power of the branches.

The objective function of the DSO which minimizes the total cost over the system is defined in (\ref{equ.6}). Equations (\ref{equ.7}) and (\ref{equ.8}) define active and reactive power balances, respectively. The generating power of each DDG is limited by (\ref{equ.9}). The power consumption by the demand response is limited with respect to the maximum value in (\ref{equ.10}). The voltage of each branch is defined in (\ref{equ.11}) and is limited with respect to the allowed voltage range in (\ref{equ.12}). The active and reactive flow of each branch is limited in (\ref{equ.13}) and (\ref{equ.14}), respectively. %More details about the formulation of this DSO can be found in our prior work \cite{mousavi2021dso}.

\bibliographystyle{IEEEtran}
\bibliography{Refs}

\end{document}